\documentclass[twocolumn,twocolappendix]{aastex631}
\usepackage{amsmath,amstext,natbib}
\usepackage{color,soul}
 
\shorttitle{}
\shortauthors{Su et al.}

\begin{document}

\title{RZ Piscium Hosts a Compact and Highly Perturbed Debris Disk}

\author[0000-0002-3532-5580]{Kate Y.~L.~Su}
\affiliation{Steward Observatory, University of Arizona, 933 N Cherry Avenue, Tucson, AZ 85721--0065, USA}

\author[0000-0001-6831-7547 ]{Grant M. Kennedy}
\affiliation{Department of Physics, University of Warwick, Gibbet Hill Road, Coventry CV4 7AL, UK}
\affiliation{Centre for Exoplanets and Habitability, University of Warwick, Gibbet Hill Road, Coventry CV4 7AL, UK}

\author[0000-0003-2303-6519]{G. H. Rieke}
\affiliation{Steward Observatory, University of Arizona, 933 N Cherry Avenue, Tucson, AZ 85721--0065, USA}
\affiliation{Lunar and Planetary Laboratory, The University of Arizona, 1629 E. University Boulevard, Tucson, AZ 85721--0065, USA}

\author[0000-0002-4803-6200]{A. Meredith Hughes}
\affiliation{Astronomy Department and Van Vleck Observatory, Wesleyan University, 96 Foss Hill Drive, Middletown, CT 06459, USA}

\author[0000-0002-1511-310X]{Yu-Chia Lin}
\affiliation{Steward Observatory, University of Arizona, 933 N Cherry Avenue, Tucson, AZ 85721--0065, USA}
\affiliation{Department of Physics, University of Arizona, Tucson, Arizona 85721, USA}

\author{Jamar Kittling}
\affiliation{Astronomy Department and Van Vleck Observatory, Wesleyan University, 96 Foss Hill Drive, Middletown, CT 06459, USA}

\author[0000-0003-4393-9520]{Alan P. Jackson}
\affiliation{School of Earth and Space Exploration, Arizona State University, 550 E. Tyler Mall, Tempe, Arizona, 85287, USA }
\affiliation{Department of Physics, Astronomy and Geosciences, Towson University, Maryland, 8000 York Road
Towson, MD 21252, USA}

\author[0000-0002-4989-6253]{Ramya M. Anche}
\affiliation{Steward Observatory, University of Arizona, 933 N Cherry Avenue, Tucson, AZ 85721--0065, USA}

\author[0000-0003-2300-2626]{Hauyu Baobab Liu}
\affiliation{3 Physics Department, National Sun Yat-Sen University, No. 70, Lien-Hai Road, Kaohsiung City 80424, Taiwan, Republic of China}
\affiliation{Institute of Astronomy and Astrophysics, Academia Sinica, 11F of Astronomy-Mathematics Building, AS/NTU No.1, Sec. 4, Roosevelt Road, Taipei 10617,
Taiwan, Republic of China}

\correspondingauthor{Kate Su}
\email{ksu@as.arizona.edu}

\begin{abstract}

RZ Piscium (RZ Psc) is well-known in the variable star field because of its numerous, irregular optical dips in the past five decades, but the nature of the system is heavily debated in the literature. We present multiyear infrared monitoring data from {\it Spitzer} and {\it WISE} to track the activities of the inner debris production, revealing stochastic infrared variability as short as weekly timescales that is consistent with destroying a 90-km-size asteroid every year. ALMA 1.3 mm data combined with spectral energy distribution modeling show that the disk is compact ($\sim$0.1--13 au radially) and lacks cold gas. The disk is found to be highly inclined and has a significant vertical scale height. These observations confirm that RZ Psc hosts a close to edge-on, highly perturbed debris disk possibly due to migration of recently formed giant planets which might be triggered by the low-mass companion RZ Psc B if the planets formed well beyond the snowlines.

\end{abstract}


\section{Introduction} 
\label{sec:intro}

\begin{figure}
    \centering
    \includegraphics[width=\linewidth]{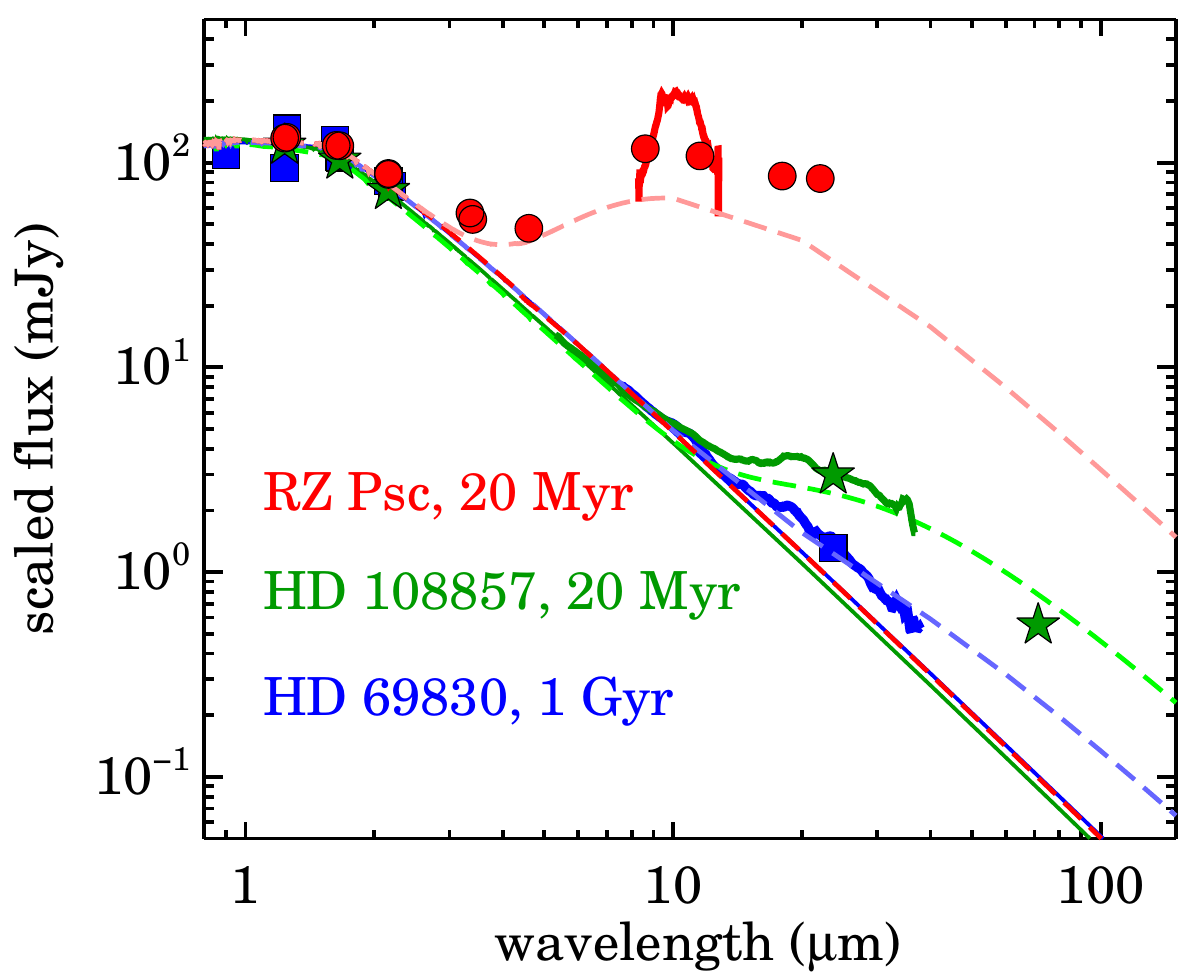}
    \caption{SED comparison for three different types of debris disks around solar-like stars: RZ Psc (red, extreme debris disk), HIP 61049 (green, typical young debris disk) and HD 69830 (blue, highly stirred warm belt) with the latter two normalized to the distance of RZ Psc. Photometry points are shown as various symbols, mid-infrared spectra as the solid-line, and model disk emission as the dashed line. The amount of infrared excess emission in EDDs is exceptionally large compared to typical debris disks.}
    \label{fig:sed_edd}
\end{figure}

Circumstellar disks of gas and dust are expected to form as a natural consequence of the star formation process, and they are sites that allow astronomers to study how planetary systems form and evolve. In the early gas-rich protoplanetary disk stage, these disks set the initial conditions for planet formation \citep{miotello22_ppvii,manara22_ppvii}; in the later gas-poor debris phase, these disks bear the imprints of planets and highlight the pathways of their formation and migration history \citep{wyatt08, krivov10,matthews14,hughes18}. A particularly interesting sub-class of debris disks is called ``extreme debris disks" (EDDs) generally found around young stars that show an exceptionally large amount of warm dust in the terrestrial zone (see Figure \ref{fig:sed_edd} for an example). Because these EDDs are mostly found around stars younger than $\sim$200 Myr, corresponding to the end of terrestrial planet accretion \citep{chambers13,quintana16}, the large amount of warm dust is commonly interpreted as the product of terrestrial planet formation \citep{melis10,meng14}.  The terrestrial planet formation link for Gyr-old systems such as HD 69830 and BD+20 307 is less clear because the extreme dustiness could also be due to (1) intense dynamical stirring via the presence of planets for the first case \citep{lovis06} or binarity for the latter case \citep{zuckerman08}, and (2) the consequence of unstable moons \citep{hansen23}. 

More than half of the EDDs are known to show photometry variability either in their total infrared outputs through {\it Spitzer} and {\it WISE}\ monitoring \citep{meng14,meng15,su19,su20,rieke21_v488per,moor21,moor22_tyc4209} or by irregular dips through routine all-sky optical surveys \citep{gaidos19_hd240779,powell21_tic400799224,melis21}. 
The characteristics of the dips are best described as star-size dust clumps, presumably generated by large collisions within a few au, passing in front of the stars \citep{dewit13_rzpsc}. Because infrared emission directly traces the production and dissipation of freshly generated dust, the combination of infrared variability and optical dips provides a powerful means to extract the details of these large collisions such as the location and the minimum size and mass of bodies involved as illustrated in the HD 166191 system \citep{su22_hd166}. 

The interpretation of the EDDs around young stars as a late stage of terrestrial planet formation is not unique. An alternative possibility is that the high dust levels are the result of transient dynamical clearing of planetesimal regions that might be analogous to the Late Heavy Bombardment in our solar system. Growing evidence has shown that the instability that shaped the reconfiguration of the giant planets in our solar system happened earlier than previously thought \citep{nesvorny18,clement19,desousa20} and was likely triggered by the dispersal of the gas disk \citep{liu22}. In addition to perturbation by newly formed giant planets, external excitements from nearly low-mass companions could also be the cause of instability. In such a scenario, the planetesimal disk might be truncated to a small radius with a significant scale height, a stage that corresponds remarkably well with the observed properties of the RZ Piscium (RZ Psc) system, the focus of this paper. 

RZ Psc is a popular, solar-like target among the amateur community \citep{aavso10_rzpsc} because it is known to show sharp irregular, optical dips over the past five decades  \citep{zaitseva78_rzpsc_dips,grinin10_rzpsc,dewit13_rzpsc,kennedy17_rzpsc}. The age of RZ Psc is heavily debated in the literature, ranging from a few Myr-old young system \citep{grinin10_rzpsc} to an evolved giant star \citep{kaminskii00_rzpsc_evolvedstar}. The most recent comprehensive study by \citet{punzi18_rzpsc}, including X-ray and multi-epoch optical, high-resolution spectroscopic observations, suggests an age of 30--50 Myr and reveals accretion activity from a reservoir of circumstellar gas. Compared to typical accreting protoplanetary disks, the gas accretion rate is low ($\dot M \lesssim 7\times 10^{-12} {\rm M_\odot\ yr^{-1}}$) as estimated from an empirical magnetospheric accretion model \citep{potravnov17_rzpsc}. Optical spectroscopic monitoring shows that the accretion rate is variable without known periodicity, and can reach one order of magnitude higher during outbursts \citep{dmitriev23_rzpsc}.  
The Galactic motion of RZ Psc calculated with new Gaia DR2 astrometric data suggests a possible membership in the Cas-Tau OB association with an age of 20$^{-5}_{+3}$ Myr \citep{potravnov19_rzpsc}. $Gaia$ EDR3 gives a distance of 184.1$\pm$0.9 pc to RZ Psc \citep{gaia16,gaia_edr3}, which is $\sim$5\% closer than the value used in earlier studies. We adopt this distance for all the derived quantities in this paper.

Although RZ Psc has all the attributions of the so-called UXOR variability commonly found among pre-main sequence stars since the 1980s, the infrared excess indicative of a circumstellar dust disk was not identified until 2013 using all-sky {\it WISE}\ and {\it AKARI}\ data \citep{dewit13_rzpsc}. The broad-band excess Spectral Energy Distribution (SED) is well represented by blackbody emission of $\sim$500 K with an infrared fractional dust luminosity of $\sim$7$\times 10^{-2}$; follow-up N-band spectroscopy also shows a prominent 10 $\mu$m feature (Figure \ref{fig:sed_edd}, \citealt{kennedy17_rzpsc}). There is no sign of cold dust (debris outside tens of au), although its presence can not be ruled out due to lack of far-infrared/millimeter observations. High contrast, ground-based imaging obtained with VLT/SPHERE recently shows that RZ Psc hosts a 0.12 M$_{\sun}$ companion at a projected separation of 22 au, and concludes that the disk (source of the infrared excess) orbits the primary star \citep{kennedy20_mnras_496_75_RZpsc_companion}. 
The low-mass component is expected to have significant impact on the disk itself such as truncating the disk outer edge and heavily stirring the remaining planetesimals. The latter is  consistent with the suggestion of a dynamically active asteroid belt revealed as numerous optical occultation events as the products of collisions \citep{dewit13_rzpsc}. 

To further shed light into the nature of the RZ Psc disk, we present multiyear monitoring data from {\it Spitzer} and {\it WISE}\ to probe the short- (weekly) and long-term (monthly to yearly) infrared variability. Millimeter observations from SMA and ALMA are also presented to assess the cold disk properties. Section \ref{sec:data} describes the data and basic reduction used in this paper. Further analyses extracted from the data are presented in Section \ref{sec:analysis} including (1) the disk properties derived from the ALMA image and SED model, indicative of a compact disk with large scale height, and (2) the general implications of the intense activity in the inner zone of the planetary system from the {\it Spitzer} monitoring. In Section \ref{sec:discussion}, we discuss the origin of the activity observed in the infrared, the potential sources of the gas reservoir, and the mechanisms that create the large disk scale height by linking with the intense dynamical activity revealed by photometric monitoring.  A short conclusion is given in Section \ref{sec:conclusion}.

\begin{table}
\begin{center}
\tablewidth{0pc}
\caption{Parameters and references for RZ Psc \label{tab:stellar}} 
\vspace{-0.2cm}
\begin{tabular}{lcl}
\hline
\hline
\multicolumn{3}{l}{Adopted Stellar Parameters} \\
\hline 
  & Adopted value & Reference \\
Luminosity $L_{\ast}$  & 0.8 L$_{\sun}$  & 1 \\
Radius $R_{\ast}$  & 1.0 R$_{\sun}$  & 1 \\
Temperature $T_{\ast}$  & 5500 K  & 1, 4, 5 \\
Mass $M_{\ast}$  & 1.1 M$_{\sun}$  & 5 \\
Distance & 184.1$\pm$0.9 pc & 6 \\
Extinction$^c$ $A_V$ & 0.18 mag & 5 \\
Age  &  $\sim$30 Myr  & 4, 5 \\
\hline
\hline
\multicolumn{3}{l}{ALMA 1.3 mm Derived Disk Parameters$^a$} \\
\hline
$F_{\rm 1.3mm}$& 42$\pm$8 $\mu$Jy &  \\
Disk peak radius $R_p$ & 6 au &  \\
Disk outer radius $R_{\rm out}$ & $\lesssim$13 au &  \\
Disk inclination $i$ & 83\arcdeg$^{+13\arcdeg}_{-34\arcdeg}$ &  \\
Disk P.A. & --16\arcdeg$^{+44\arcdeg}_{-20\arcdeg}$ &  \\
Dust mass & 6.4--40 $\times 10^{-3} {\rm M_\oplus}$ &  \\
CO gas mass & $<$1.0 $\times 10^{-5} {\rm M_\oplus}$ &  \\
\hline
\hline
\multicolumn{3}{l}{SED Derived Disk Parameters$^b$} \\
\hline
  & best-fit & search range \\ 
Inclination $i$  & $\sim$75\arcdeg & 40--90\arcdeg \\
Outer radius $r_{out}$ & 12$^{+1}_{-2}$ au & 1--15 au \\
Scale height $H$ & 11$^{+2}_{-2}$ au & 0.1--15 au \\
Disk mass  & 2.5$\times 10^{-3}$ ${\rm M_\oplus}$ & 4--400$\times 10^{-4}$ ${\rm M_\oplus}$ \\
\hline
\end{tabular}
\end{center}
\tablenotetext{}{$^a$ details see Section \ref{sec:alma_intep}. \\
$^b$ details see Section \ref{sec:sed_mod}. \\
$^c$ interstellar only. \\
References: [1] this work; [2] \citet{schneider13}; [3] \citet{kennedy14}; [4] \citet{punzi18_rzpsc}; [5] \citet{potravnov18}; [6] {\it Gaia} EDR3, \citet{gaia_edr3}. } 
\end{table} 

\section{Observations and Data Reductions} 
\label{sec:data}

\begin{figure*}
    \includegraphics[width=\linewidth]{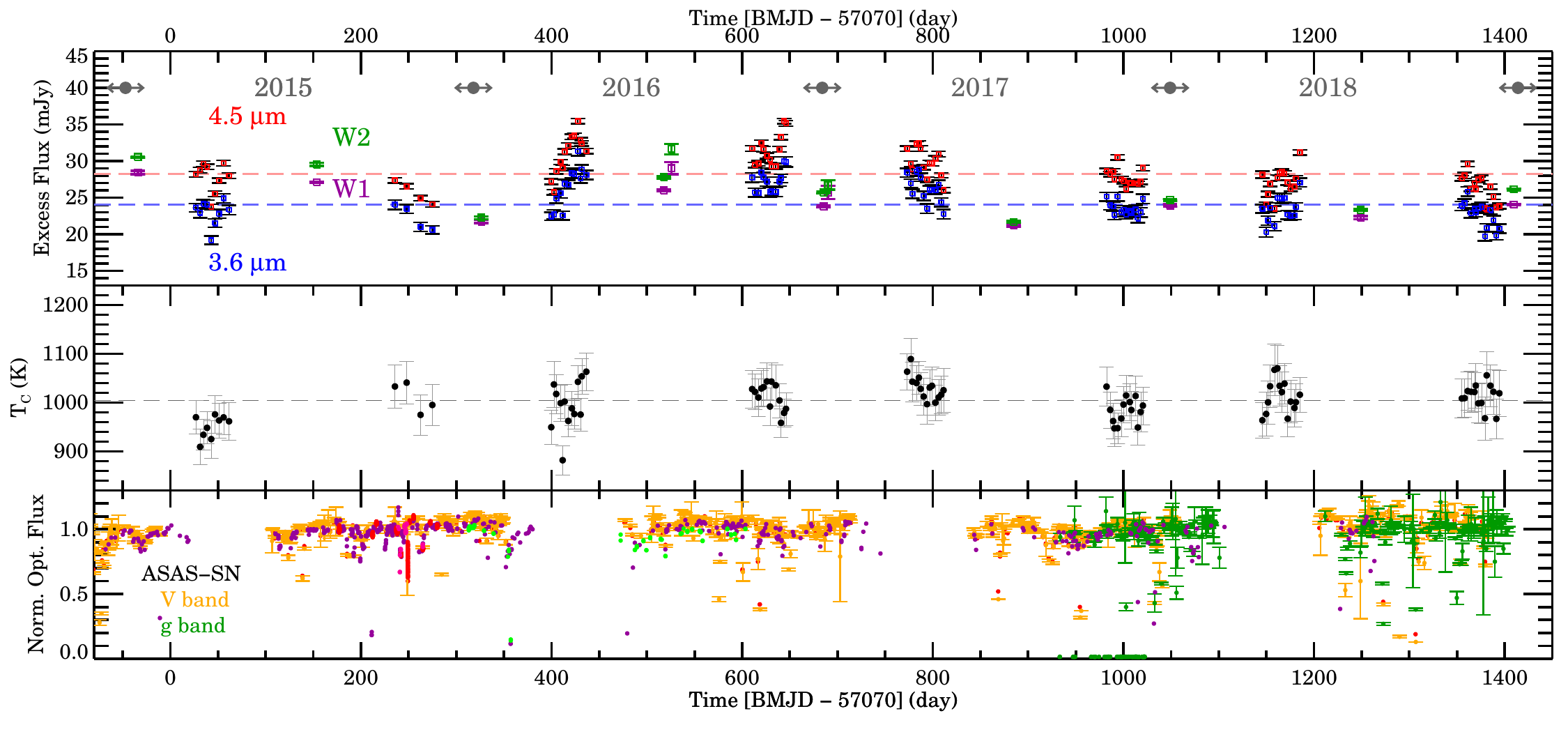}
    \caption{{\it Spitzer} monitoring data of the RZ Psc system: top for the 3.6 and 4.5 $\mu$m excess fluxes, middle for its corresponding color temperature, and bottom for the optical light curves. The arrow-like symbols near the top of the first panel mark the year boundaries of the data. Horizontal dashed lines in the upper two panels are the median value of the derived quantity. Optical light curves include ASAS-SN $V$- (orange) and $g$-band (green) data, and AAVSO $V$-band (various colors) data (for details see Section \ref{sec:opt_obs}).}
    \label{fig:irac_rzpsc}
\end{figure*}

\subsection{Warm {\it Spitzer}/IRAC} 
\label{sec:irac_obs}

Warm {\it Spitzer} monitoring data were obtained under two GO programs PID 11093 and 13014 (PI Su) with observations covering from 2015 March to 2018 December. RZ Psc has two {\it Spitzer} visibility windows per year, each a total of $\sim$36 days in length. During the first year of the monitoring, cadences of $\sim$5$\pm$1 and $\sim$10$\pm$2 days were used to search for weekly variation due to material as close as 0.1 au from the star. We later used a finer cadence of 3$\pm$1 days in the 2016--2018 observations. A total of 95 sets of observations at both 3.6 and 4.5 $\mu$m bands was obtained. We used a frame time of 2.0 s with 15 cycling dithers (i.e., 15 frames per Astronomical Observation Request (AOR)) to minimize the intra-pixel sensitivity variations of the detector \citep{reach05_irac} at both bands, achieving a signal-to-noise ratio of $\gtrsim$180 in single-frame photometry. These data were first processed with IRAC pipeline S19.2.0 by the {\it Spitzer} Science Center.  We performed aperture photometry on each of the data sets, following the procedure outlined in \citet{su19}. The final weighted-average photometry is given in Table \ref{tab:irac}. We also determined the instrumental photometry repeatability by monitoring two isolated sources in the field of view. A $<$1\% rms in the measured photometry was found for nonvarying sources in these data.  

Using the stellar parameters listed in Table \ref{tab:stellar}, we estimated the stellar photosphere in the two IRAC bands to be 34 and 22 mJy with a typical uncertainty of 1.5\% limited by the accuracy of the optical and near-infrared photometry. After subtracting the stellar contribution by assuming the star is stable at these infrared wavelengths, the infrared excesses (i.e., the disk fluxes) and associated color temperatures are shown in Figure \ref{fig:irac_rzpsc} over the four-year span in the units of Barycentric Modified Julian Date (BMJD). Overall, the infrared output of the system at 3.6 and 4.5 $\mu$m shows stochastic variations with a 50--60\% maximum peak-to-peak flux difference. No significant periodicity was found using several periodogram tools, except for spurious periods associated with the sampling windows. The median values of the excess fluxes are 24.0 mJy and 28.3 mJy at 3.6 $\mu$m and 4.5 $\mu$m, respectively (shown as horizontal dashed lines in the upper panel of Figure \ref{fig:irac_rzpsc}). Because the star dominates the noise in the measured photometry, the typical uncertainty in the two-band color temperatures is $\sim$40--50 K with an average of 1000 K (the horizontal dashed line in the middle panel of Figure \ref{fig:irac_rzpsc}). We note that the color temperature derived from the two shortest IRAC bands is not necessarily equal to the dust temperature in the system. Under optically thin conditions, the derived color temperature might reflect the true dust temperature, but the absolute numbers also depend on the assumed photospheric fluxes that are subtracted off from the measurements. A $\sim$10\% change in the photospheric fluxes could result in changes to the derived color temperature on the order of $\sim$100 K in the case of RZ Psc.

\subsection{{\it WISE/NEOWISE} and Optical Photometry}
\label{sec:opt_obs}

\begin{figure}
    \centering
    \includegraphics[width=\linewidth]{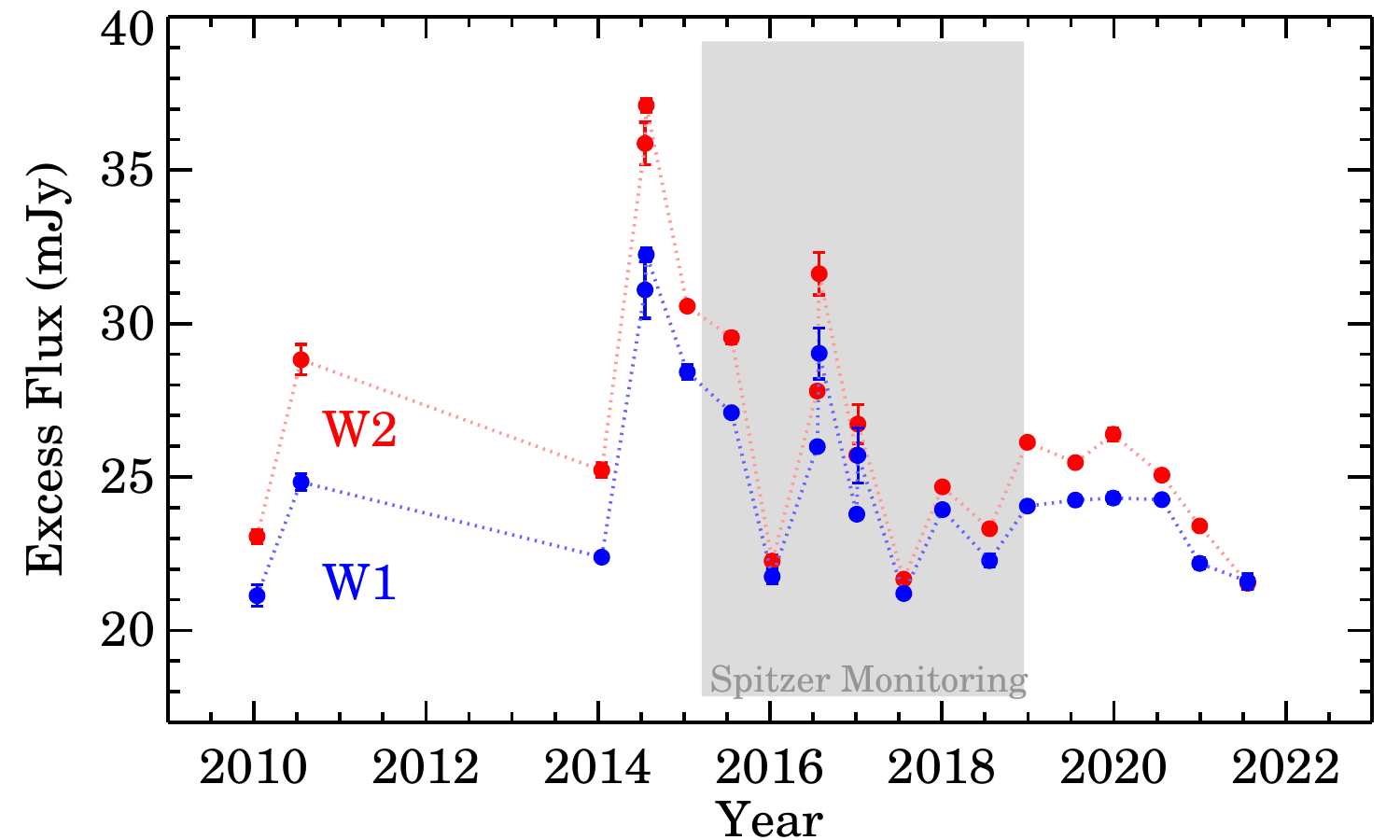}
    \caption{{\it WISE/NEOWISE} W1 and W2 data for the RZ Psc system where the stellar contribution has been subtracted off. The peak infrared flux occurred at mid-2014, prior to the {\it Spitzer} monitoring period (shaded region).}
    \label{fig:wise_longterm}
\end{figure}

{\it NEOWISE} provides all-sky W1- and W2-band photometry every 6 months since 2014 \citep{neowise_ref}, a time-domain resource for variability study. We extracted the single-exposure photometry from IRSA to fill the gaps between {\it Spitzer} visibility windows and to assess the long-term variability including data from the cryogenic {\it WISE} mission. For easy viewing, we binned the single-exposure data using a cadence of 3 days and plotted\footnote{Due to the slight different transmission between {\it Spitzer} and {\it WISE} filters, an offset of -2 and -0.5 mJy, for W1 and W2 respectively, were applied to the {\it WISE/NEOWISE} photometry.} them in Figure \ref{fig:irac_rzpsc} when overlapping with the {\it Spitzer} data. Because the {\it WISE/NEOWISE} mission gives more than 10 years of W1 and W2 photometry, the multiyear WISE W1/W2 photometry is shown in Figure \ref{fig:wise_longterm}. {\it WISE/NEOWISE} data reveal that the overall infrared excess peaked in mid-2014 and gradually declined to the end of 2021, back to the 2010 level.

We also made use of the publicly available optical photometry from ASAS-SN Sky Patrol (\url{https://asas-sn.osu.edu/}, \citealt{shappee14, kochanek17}) and AAVSO. For ASAS-SN, we binned/combined and normalized the $V-$ and $g-$band data following the procedure outlined by \citet{su22_hd166}. In the AAVSO database, there were lots of data from different observers with slightly different filters. We selected data from five different observers who used $V$-band filters and have enough measurements to establish ``zero" points just like treating different cameras used in the ASAS-SN survey. All the normalized optical photometry is shown in the bottom panel of Figure \ref{fig:irac_rzpsc}.

\begin{figure*}
    \centering
    \includegraphics[width=\linewidth]{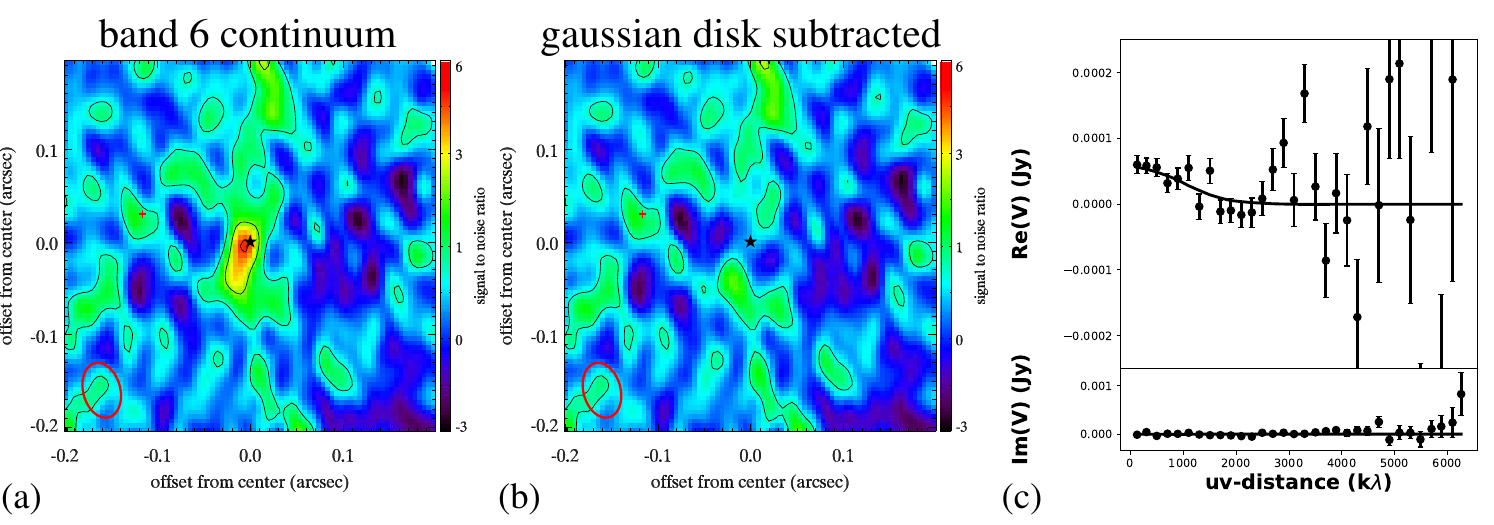}
    \caption{(a) A naturally weighted ALMA image of the 1.3 mm continuum observation centered at RZ Psc (star symbol). Contours are [1, 3, 5]$\times\sigma$, where $\sigma$= the rms noise = 8 $\mu$Jy beam$^{-1}$. The synthesized beam dimensions, represented by the red ellipse in the lower left corner, are 0\farcs06$\times$0\farcs04 with a position angle of 14.8\arcdeg. The position of RZ Psc B \citep{kennedy20_mnras_496_75_RZpsc_companion} is marked by the red plus for reference. The middle and right panels are the corresponding residual maps after the subtraction of a Gaussian-disk model in (b) and the associated uv-data in (c) (Details see Section \ref{sec:alma_intep}).  }
    \label{fig:alma}
\end{figure*}

\subsection{SMA Observation} 
\label{sec:sma_obs}

RZ Psc was observed with the SMA at 225 GHz in 2015. Seven antennae in both compact (on September 8) and extended (on September 19) array configurations were used, covering a {\it uv} distance range of $\sim$10--170 $k\lambda$. The atmospheric opacity ($\tau_{\mbox{\tiny 225 GHz}}$) was $\sim$0.3 and $\sim$0.07 during these two tracks of observations, respectively. The observations were carried out in single receiver mode, which covered the 212.7--220.7 GHz observing frequency in the lower sideband, and the 228.8-236.8 GHz observing frequency in the upper sideband. Correlations were performed by the ASIC correlator and the SMA Wideband Astronomical ROACH2 Machine (SWARM). Uranus and Quasar 3C84 were used for absolute flux and passband calibration. Standard SMA data reduction packages \citep{Qi2003,Sault1995} were used to calibrate and generate the final continuum and line maps. Using natural weighting  yielded a synthesized beam of 1$\farcs$3 $\times$ 1$\farcs$0 at a position angle (P.A.)=--81$^{\circ}$, and an root-mean-square (rms) noise level of 0.6 mJy\,beam$^{-1}$. RZ Psc was not detected in the continuum map. We also imaged the CO 2-1 line (rest frequency: 230.538 GHz) with 1.2 km\,s$^{-1}$ channel width, which achieved a 1$\farcs$2 $\times$ 0$\farcs$97 (P.A.=--80$^{\circ}$) synthesized beam, and an rms noise level of 44 mJy\,beam$^{-1}$. Similarly, no line was detected. Because these non-detection limits are much less stringent than the ALMA observation (next Section), we will not discuss the SMA data further.

\subsection{ALMA Observations} 
\label{sec:alma_obs}

ALMA observations were taken on 2021 August 22 and 23 under the project 2019.1.01701.S (PI: K. Su). Two observing blocks each with 36 min of on-source integration were obtained using the Band 6 receiver and 49/44 12m antennas with baselines ranging from 47 m to 11.6 km under a mean precipitable water vapor (PWV) column of 0.3 mm. Three spectral windows were devoted to the continuum at 245 GHz with a 2 GHz bandwidth, while one was centered on the CO(2-1) 231 GHz line with 1920 channels (0.31 km s$^{-1}$). J0238+1636 was the main calibrator for pointing, atmosphere, bandpass, phase and flux calibration. Calibrated data products were produced and provided by the North American ALMA Science Center using CASA version 6.1.1. Because RZ Psc is a young low-mass star, we also checked for potential mm variability due to chromospheric and/or accretion-related events. To search for point source emission, we plotted the baseline-averaged visibility amplitudes as a function of time for each date of observation.  Since a point source has a constant amplitude as a function of baseline length, baseline averaging should reveal any point source variability as a function of time within the data set, yet all points were consistent with a constant flux density using the 2 and 30 s average data. Since the known mm flares have all taken place on timescales of minutes or less \citep{macgregor20}, there is no mm flaring event during our ALMA observation.

We imaged the continuum by combining all four spectral windows using natural weighting, resulting in an rms of 8 $\mu$Jy\,beam$^{-1}$ with a beam size of 0\farcs06$\times$0\farcs04 at a P.A. of 14.8\arcdeg. There appears to be one bright source detected near the expected RZ Psc position with a peak flux of 41.6 $\mu$Jy\,beam$^{-1}$ surrounded by a faint, elongated extended structure (Figure \ref{fig:alma}). The source extension (defined by the 3$\sigma$ contour) on the synthesized image is 0\farcs20$\times$0\farcs07 at a P.A. of $-$13\arcdeg. Given the source's location (north of $\sim$28\arcdeg), the synthesized beam is extremely elongated, but along a different P.A. from that of the extended structure. The comparison between the source extension and the synthesized beam, particularly the P.A. (15\arcdeg\ vs.\ $-$13\arcdeg), suggests the source is spatially resolved. We will discuss the fidelity of the extended structure in Section \ref{sec:alma_intep}. 

Using the $Gaia$ DR2 proper motion values (pmRA=27.544$\pm$0.028 mas/yr and pmDEC=$-$12.521$\pm$0.019 mas/yr) for RZ Psc, there is a small offset ($-$7 mas in both RA and Dec) between the detected and expected positions. The pointing accuracy is $\sim$10 mas with the synthesized beam and signal-to-noise above, i.e., this small offset is insignificant. We conclude that the emission detected in the 1.3 mm map is from the RZ Psc system, and most likely the dust emission around the star because the stellar photosphere is expected to be at 0.3 $\mu$Jy. While there is a suggestive peak that coincides with the position of the companion RZ Psc B, it is only at the 1.5 $\sigma$ level and is therefore entirely consistent with expected noise properties. At the distance of 184 pc and the $H$/$K_{\rm S}$ magnitudes for the M-dwarf companion \citep{kennedy20_mnras_496_75_RZpsc_companion}, the companion is at least 10 times fainter than RZ Psc at 1.3 mm. Although we cannot completely rule out the possibility of excess emission around the companion, it is more likely just a noise spike in the data.

For the CO(2-1) observation, we concentrated on the spectral windows that contain the CO 230.538 GHz rest frequency in the two days of data by averaging in 1s time intervals to avoid time smearing. We used the $cvel$ command to convert the velocities from topocentric (TOPO) to kinematic local standard of rest (LSRK). We then converted the measured heliocentric velocity (2 km s$^{-1}$, \citealt{punzi18_rzpsc}) to LSRK velocity of 0.91 km s$^{-1}$.  After subtracting the continuum with the $uvcontsub$ task, we used $tclean$ to generate channel maps centered on the rest frequency, and found no significant emission on a channel-by-channel basis. The zeroth moment map of CO was then constructed by integrating channels within $\pm$10 km s$^{-1}$ of the star's LSRK velocity. Assuming the disk is unresolved, we determine the 3$\sigma$ upper limit of CO (2-1) flux is 1.03$\times10^{-2}$ Jy~km~s$^{-1}$.  We will further discuss the upper limit for the CO gas in Section \ref{sec:alma_intep}.

\section{Analysis}
\label{sec:analysis}

\subsection{Disk Parameters from the ALMA Observation}
\label{sec:alma_intep}

To further evaluate the significance of an extended structure in the ALMA data and constrain the disk extent, we fit the observations in the visibility domain with simple parametric models. We adopt two models that are centered on the star: (1) the disk emission has a Gaussian-like profile, and (2) the disk emission comes from a narrow (unresolved) ring. The common parameters for both models are the P.A. of the disk major axis, the disk inclination, and total flux, and the different ones are in terms of disk size where the former one only constrains the Gaussian core size $\sigma$, which is related to the Full-Width-Half-Maximum (FWHM) of the disk emission (FWHM = 2.354 $\sigma$), and the latter estimates the peak radius ($r_0$) of the narrow ring. For the Gaussian-like model, we found the Gaussian $\sigma$=0\farcs03$^{+0\farcs03}_{-0\farcs02}$, i.e., the best-fit FWHM of the disk emission is 0\farcs07 (just slightly larger than the beam) and at a P.A. of $-$13\arcdeg, consistent with the 3$\sigma$ contour. The residual map and associated uv-data are also shown in Figure \ref{fig:alma}. For the narrow-ring model, we found $r_0$=0\farcs03$^{+0\farcs04}_{-0\farcs02}$, similar to the Gaussian-disk model. The disk is really compact with the peak radius $\sim$0\farcs03 (6 au at 184 pc), and only slightly larger than the beam with the maximum radius less than 0\farcs07 (13 au). Given the compactness of the disk, the other geometric parameters like the disk inclination and P.A. are less constrained. By combining the two models, we estimate the disk inclination angle is  83\arcdeg$^{+13\arcdeg}_{-34\arcdeg}$, and at a P.A. of $-$16\arcdeg$^{+44\arcdeg}_{-20\arcdeg}$ with a total flux density of 42$\pm$8 $\mu$Jy at 1.3 mm. The large uncertainties associated with the derived parameters corroborate the fact that the disk is only marginally resolved (i.e., mm emission peaks at 6 au with a maximum radius less than 13 au) and the inclination and P.A. are not well constrained. Finally, the disk is unresolved vertically, and we can place an upper limit on the mm disk scale height $<$28 au (3 $\sigma$ at 184 pc). 

We can convert the 1.3 mm flux to a total dust mass (including grains up to mm--cm sizes) by assuming a dust opacity of 2.3 g~cm$^{-2}$ \citep{beckwith90} and a dust temperature of 20--100 K\footnote{The thermal equivalent dust temperature is $\sim$100 K at 10 au from RZ Psc assuming blackbody-like grains. Therefore, we set the higher end temperature to be 100 K rather than the higher inferred temperatures from the {\it Spitzer} observation (i.e., 500 K corresponds to a few tenths of an au under the same conditions.)}, which corresponds to 6.4--40$\times10^{-3}$ M$_\oplus$. Using the 3-$\sigma$ upper limit of CO (2-1) flux (Section \ref{sec:alma_obs}) and an excitation temperature of 40 K, the corresponding upper limit of the cold gas mass is 1.0$\times 10^{-5}$ M$_{\oplus}$. The ALMA observation indicates a very low gas-to-dust mass ratio $<10^{-3}$ in the system, confirming that RZ Psc has evolved out of the gas-rich protoplanetary stage and into the debris phase.

\subsection{SED model}
\label{sec:sed_mod}

To test whether the observed broad-band SED from infrared to millimeter is largely consistent with disk parameters extracted from the ALMA observations mainly the disk outer radius), we employ SED modeling of a single disk component with the publicly available radiation transfer code, MCFOST \citep{mcforst}. For simplicity, we assume an azimuthally symmetric, wedge-like (i.e., a fixed opening angle) disk and fix the majority of the parameters in the model, and mainly explore four disk parameters: inner and outer disk radii, disk scale height and total dust mass. The disk surface density is assumed to follow a $r^{-3/4}$ distribution where $r$ is the radius. Because there is no/very little gas detected by ALMA, we set the disk flaring index to 1, i.e., the disk scale height, $H$, as $H = H_o (r/r_o) $ where $r_o =$ 10 au. The star with the parameters listed in Table \ref{tab:stellar} is the only heating source. We also fix the grain properties by adopting Astrosilicates as the composition in a power-law size distribution $n(a)\sim a^{-3.5}$ (where $a$ is the grain radius) from $a_{min}=$ 0.1 $\mu$m (to match the prominent 10 $\mu$m silicate feature) to $a_{max}=$ 1 cm (to account for the millimeter emission). An initial search on a wide parameter space suggests that the inner radius of the disk is close to 0.1 au in order to account for the emission between 2--5 $\mu$m. The temperature at that radius is close to the dust sublimation temperature for sub-$\mu$m silicate-like grains. We, therefore, also fix $r_{in}=$ 0.1 au. 

\begin{figure}
    \centering
    \includegraphics[width=\linewidth]{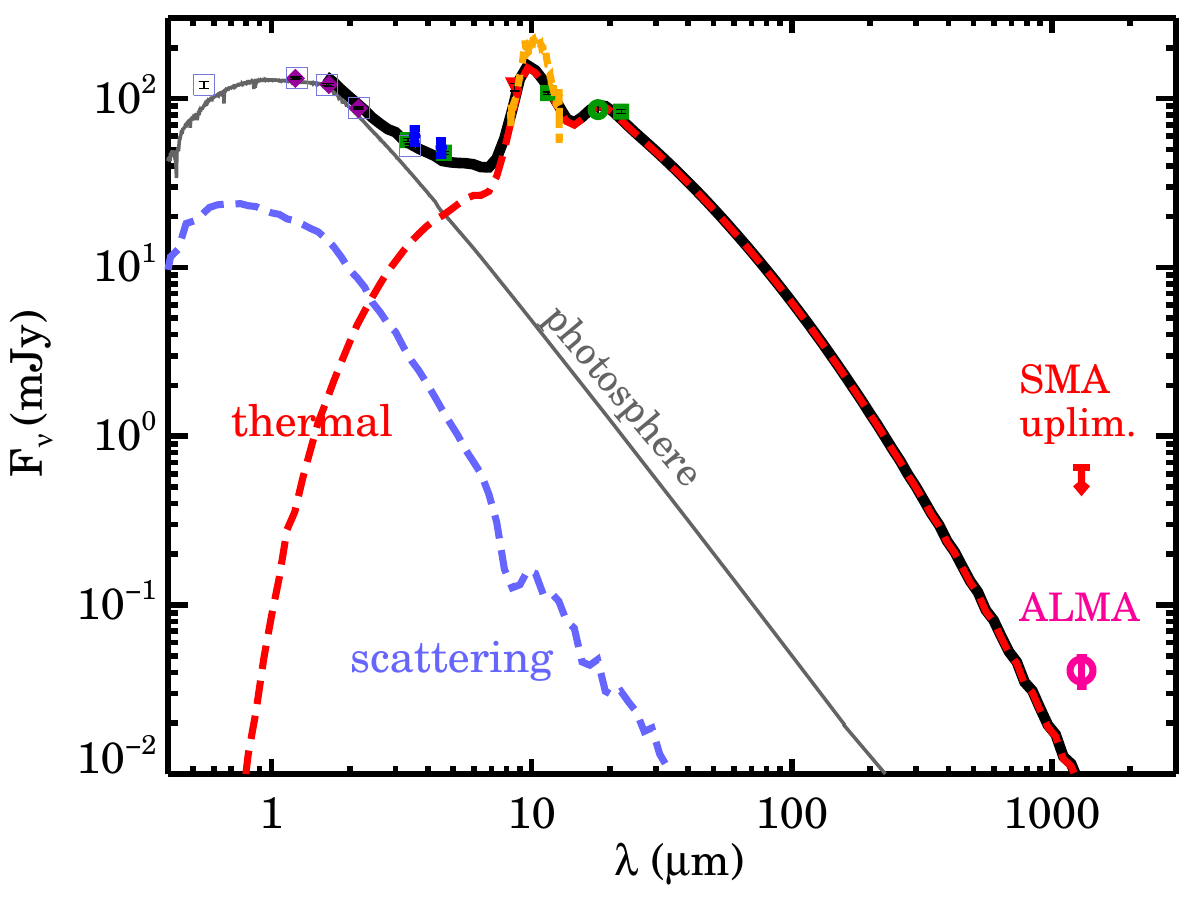}
    \caption{SED models of the RZ Psc system including the star and a single-component disk (see Section \ref{sec:sed_mod} for details). 
    Various symbols are publicly available broad-band photometry points from Simbad, 2MASS \citep{2mass_catalog}, ALLWISE \citep{cutri14}, and {\it AKARI} \citep{akari_irc_catalog}, and 
    the yellow dashed line is the $N$-band grism spectrum from \citet{kennedy17_rzpsc}.}
    \label{fig:sed_diskmodel}
\end{figure}

We employ the Python package $emcee$ \citep{emcee} to derive best-fit values for the disk outer radius, scale height and total dust mass. The results are shown in Table \ref{tab:stellar} and the best-fit SED is shown in Figure \ref{fig:sed_diskmodel}. Interestingly, the model millimeter flux is $\sim$3 times lower than the ALMA measurement with  $r_{out}$ = $12^{+1}_{-2}$ au. This is why the SED derived dust mass is much lower than the direct mass conversion from the millimeter flux (Table \ref{tab:stellar}). A larger outer disk radius would increase the millimeter flux (i.e., the mass), but such models have too much emission in the 2--5 $\mu$m region. This suggests that the disk structure might not be as simple as we assume. For example, the disk might have multiple components (i.e., a separate cold component that emits mostly in the far-infrared and millimeter wavelengths) or the disk is not axi-symmetric (i.e., the radial temperature difference between the apses of an eccentric ring might account for the missing disk millimeter emission). A two-component model (inner and outer) would certainly fit the SED as long as the outer cold component does not contribute much of the infrared flux (i.e., either the cold component lacks sub-$\mu$m grains and/or the location of the cold component is outside of $\sim$5 au). We note that a cold component extending to 3 times larger than the ALMA size would also fit the observed SED, really illustrating the degeneracy of SED modeling. 

We also note that the favored inclination angle is $\sim$42\arcdeg\ in the SED search. Although within the possible range measured by ALMA, such models produce too much optical polarization ($\sim$5\% level at V-band), $>$5 times higher than the observed averaged value \citep{shakhovskoi03}. In fact, we find that the polarization percentage can be significantly reduced if the disk is more inclined like $i \sim$75\arcdeg\ while allowing similar fits to other photometry points (i.e., inclination does not have a big impact on the SED models within the search range because the disk is not very optically thick). Because $i \sim$75\arcdeg\ is more consistent with the values derived from the ALMA data and the optical transits, we adopt it as the best-fit value. Optical polarimetry is very sensitive to the disk inclination, therefore, polarization measurements provide much stronger constraints on the disk inclination angle. We will explore further constraints on the disk structure (inclination and the degree of asymmetry) using time-series optical polarimetric measurements in a coming paper (Anche et al.\ in prep.). 

Finally, a scale height of 12$^{+2}_{-2}$ au at the radius of 10 au indicates that the disk has a very wide opening angle, $\sim$50\arcdeg. Such a large opening angle is in stark contrast with the nominal angle (a few degrees) measured in typical debris disks. Although the SED derived disk scale height is degenerate with many other parameters (i.e., the exact quantity is uncertain), a larger than normal debris disk scale height is evident in comparison of the SEDs shown in Figure \ref{fig:sed_edd}.  Both the 20 Myr-old systems, RZ Psc (an EDD) and HIP 61049 (a typical young debris disk), show solid-state features in the mid-infrared that trace small grains in the optically thin part of the disk (usually on the surface of the disk above the midplane). The prominence of the mid-infrared features (i.e., the equivalent width of the feature) is known to be sensitive to the disk scale height in the SEDs of protoplanetary disks, which has a similar effect in the SED models of a dusty disk around white dwarfs (see Figure 4 in \citealt{ballering22}). We emphasize that the prominence of the mid-infrared features are also affected by other parameters in the model, particularly the amount of small grains. The disk scale height derived from our SED model should be taken with a grain of salt. Nonetheless, a linear size of 12 au at 184 pc would have an angular size of 0\farcs065, consistent with not being resolved by the ALMA observation.  Based on the marginally resolved ALMA map and the SED behavior, we can conclude that RZ Psc hosts a highly inclined  ($\sim$75\arcdeg), compact ($\sim$0.1--13 au) but vertically extended debris disk.
We will discuss the possible mechanisms to create this specific structure in Section \ref{sec:puffed-up}.

\subsection{Implication from Spitzer Observations}
\label{sec: spitzer_intep}

While RZ Psc has been well-known for its optical variability both in broad-band photometry and polarimetry over decades, its infrared variability was recently noticed by \citet{kennedy17_rzpsc} using initial {\it WISE} and {\it AKARI} data. It is clear that the infrared variability cannot come from the star itself because the star is not large enough to account for the infrared flux change, i.e., the changes must come from the circumstellar environment either due to (1) new debris generated in the collisions and aftermaths between large bodies such as $\gtrsim$100 km size asteroids and planetary embryos or (2) heating variation in the existing circumstellar material such as the dust clumps seen partially blocking the star. Although both are likely, the second case would make it difficult to explain the increasing infrared flux, and the source of occultation (naturally produced by the first case) requires further explanation.  For simplicity, we only consider that the infrared variability arises from the changes in the dust mass and distribution. 

In a typical debris system where its infrared emission is sustained by the collisional cascades of km-size planetesimals, the production and loss of dust debris is balanced and no short-term variation is expected. Whether a source is variable or not is subject to the detectability, largely driven by the instrument sensitivity and the frequency of sampling. Because the change in flux is directly related to the change in the total cross section of the emitting material ($\Delta \Sigma$), one can quantify the detectability of variations for a given a sensitivity. For example, a flux change of 1 mJy at 4.5 $\mu$m ($\Delta F_{\rm 4.5}$ = 1 mJy) around a solar-type star at a distance of 200 pc requires a minimum change in the surface area of $\sim$12 stellar sizes (8$\times 10^{-4}$ au$^2$) assuming a dust temperature of 600 K. $\Delta \Sigma$ is scaleable as $\frac{\Delta F_{\rm 4.5}}{1\ {\rm mJy}} [\frac{200\ {\rm pc}}{\rm d}]^2$ where $d$ is the distance, and is sensitive to the dust temperature so that the hotter the temperature the smaller the surface area. In other words, within the current infrared sensitivity limits from {\it WISE} and {\it Spitzer}, the detectable infrared variability has to come from large-scale events, such as violent impacts between large bodies, to provide enough change in the surface area. 

In addition to the infrared flux variability, the change in the color temperature derived from W1/W2 or IRAC1/IRAC2 bands can also shed light on the evolution of newly-generated debris. Under the optically thin condition, there are two mechanisms to manifest as a temperature change: one is a change in the dust location which goes as $T_d \sim r^{-0.5}$, and the other is a change in the dominant grain size which goes as $T_d \sim a^{-1/6}$. To extract the physical quantity, one needs to know the baseline level, which is difficult to establish for a constantly varying system like RZ Psc.

It is interesting to note that the {\it Spitzer} high-cadence ($\sim$3--5 days) data reveal that the debris emission (i.e., the excess flux) varies on a weekly scale, and the longest period without a large degree change is about two weeks ($\sim$4 sequential {\it Spitzer} measurements). Over the four-year {\it Spitzer} monitoring period, the highest flux of the dust emission occurred in 2016 (near the display dates of 428 and 644 in Figure \ref{fig:irac_rzpsc}), $\sim$35.5 mJy at 4.5 $\mu$m, while the lowest flux occurred near the beginning of 2015 (the display date of 43) and the end of 2018 (the display date of 1380), $\sim$23.4 mJy at 4.5 $\mu$m. The variation behavior is consistent with the long-term behavior observed in {\it WISE} (Figure \ref{fig:wise_longterm}). Compared to the {\it WISE/NEOWISE} data, the highest flux observed by {\it Spitzer} is $\sim$86\% of the peak flux in 2014, and the lowest is at a similar level ever observed (in 2010, 2016/2017).  Using the average flux as the baseline, the typical change of flux at 4.5 $\mu$m is $+$7 and $-$5 mJy for increasing and decreasing cases, respectively. 

Putting the observed infrared variability into context, one can estimate the minimum change in the emitting dust surface area (hence the minimum change in the associated dust mass) given an object's distance (184 pc for RZ Psc) and dust temperature ($\sim$1000 K from Section \ref{sec:irac_obs}). Adopting the typical flux changes observed at 4.5 $\mu$m (5--7 mJy), the change in the dust emitting surface area is (3.9--5.4)$\times10^{-4}$ au$^2$, i.e., the associated dust mass is (0.3--1.1)$\times10^{21}$ g assuming a dust density of 3.5 g~cm$^{-3}$ and a Dohnanyi-like size distribution \citep{dohnanyi69} with a minimum grain size of 0.1--0.5 $\mu$m. We note that these numbers are lower limits because of the assumption that the dust emission is optically thin, and are sensitive to the adopted dust temperature: $\sim$10\% change in the assumed dust temperature would result in $\sim$50--100\% change in the derived dust mass. A total dust mass of 1.1$\times10^{21}$g is roughly equivalent to the mass of a 90-km diameter asteroid. Such a flux change was recorded at least, if not more than, once per year. The RZ Psc system has experienced intense collisional activity, i.e., equivalent to total destruction of a 90-km asteroid every year (mass loss rate of $\sim$3.5$\times10^{13}$g~s$^{-1}$). It is unlikely such a high mass loss rate has persisted over the entire age of the system (20--50 Myr), i.e., requiring a mass equivalent to $\sim$4--9 $M_{\oplus}$. In other words, if such a mass loss rate persisted over a period of 10$^3$ (10$^6$) yr, the associated dust mass is roughly equivalent to the mass of the Moon (Earth).

\section{Discussion} 
\label{sec:discussion}

\subsection{Origin of High Collision Rate in the Inner Disk}
\label{sec:origin-variability}

ALMA 1.3 mm data combined with SED modeling show that the RZ Psc disk lacks cold gas and is composed of dust in a compact form ($\sim$0.1–13 au radially), in stark contrast to typical UXOR-type young disks. As discussed extensively by \citet{kennedy17_rzpsc}, the dust occulting events that dim the star in the optical must arise from a small fraction of the planetesimal population and lie at relatively high vertical positions above the main disk, either due to their high orbital inclinations or because they have been ejected into the inner region. The color change in the star during an occultation is similar to that expected from typical interstellar dust \citep{shakhovskoi03}, indicating that sub-$\mu$m particles play a dominant role in extincting the stellar emission. The presence of sub-$\mu$m grains is also consistent with the distinct 10 $\mu$m feature \citep{kennedy17_rzpsc}. The measurements with IRAC at 3.6 and 4.5 $\mu$m show a color temperature of $\sim$1000K, suggesting material exists inside an au. 
These observations indicate that much of the dust that dominates both the occultations and the 3--5 $\mu$m emission/variability is from grains small enough ($\lesssim 1$ $\mu$m for a solar-like system, \citealt{arnold19}) that they are ejected by radiation pressure force. This conclusion is consistent with the rapid fading of the events in the infrared discussed above and the failure of the occulting clumps to survive for a full orbital period of the inner disk \citep{kennedy17_rzpsc}. RZ Psc represents an  extreme level of chaotic variability at 3--5 $\mu$m among all known debris disks, requiring episodic dust creation. Taking these observations together, they indicate a very high level of collisional activity among the planetesimals in the inner disk, that can produce large clouds of very finely divided dust and in extreme cases eject it well above the main planetesimal disk plane. 

Can the short-term infrared variability result from gravitational stirring by the binary companion directly?  The stability of planets and planetesimals in a binary system has been explored in many studies \citep{cuntz_musielak07,nesvold16,quarles20}, and strongly depends on the properties of the binary (the mass ratio and detailed orbital parameters of the companion), as well as the starting orbital parameters of the planetesimals/planets relative to the binary orbit. The current available properties of RZ Psc B (mass and projected distance) suggest that planetesimals in the outer $\sim$10-au region might be greatly influenced by the companion through the Kozai-Lidov Effect \citep{naoz16review_kozai}, but the ones within $\lesssim$ an au are well within the stable range. This result is intuitively reasonable, since the gravitational force  from the companion star is less than 1/10,000 that from the primary star in the inner region.  A number of items of circumstantial evidence also support this conclusion. First, binary stirring is likely to have persisted for a significant fraction of the life of the star, but the mass loss rates computed in Section \ref{sec: spitzer_intep} seem unreasonably high if the perturbation has been active over half of the lifetime of the system. Second, other binary systems do not have debris systems nearly as extreme as that of RZ Psc  \citep{trilling07,yelverton19}. 

An interesting alternative possibility is that the extreme debris activity is a result of planetary migration or planet-planet scattering. RZ Psc is of an age where planetary migration and scattering are thought to have been rather common. Migration proceeds by scattering of planetesimals, which deflects the small-body population into highly eccentric and inclined orbits \citep{levision07_PPV}. During planet-planet scattering, ``the planets roam around the system on eccentric and inclined orbits for an extended period" quoted from \citet{marzari14}, creating a period of chaotic evolution. Either mechanism shifts large numbers of planetesimals onto colliding orbits and will result in large amounts of collisionally produced dust.

The fate of this dust depends on its size (and composition). Grains $\lesssim$ 1 $\mu$m in radius will be ejected via radiation pressure force \citep{arnold19}, exhibiting rapid drops in the infrared flux \citep{su19}. Grains larger than the radiation blowout size would also drift outward radially with the aid of tenuous gas \citep{kenyon16,najita23}. Larger grains will experience drag forces from the stellar wind (particularly important for young solar-like systems) and Poynting--Robertson drag. The conditions for this process around RZ Psc appear to be very similar to those for V488 Per discussed in  \citet{rieke21_v488per}. We adapt the estimate in that paper to find that 

\begin{equation}
\tau_{wind} \sim \frac{1}{\beta} \left( \frac{r_0}{\rm au} \right)^2 {\rm yr},
\end{equation}

\noindent
where $\tau_{wind}$ is the time for a grain size expressed as $\beta$ (a ratio of radiation pressure force to gravitational force) to be dragged into the star from an initial distance of $r_0$. For example, a grain with $\beta = 0.01$, corresponding to a radius up to $\sim$ 30 $\mu$m \citep{arnold19}, will be drawn from 0.50 to 0.49 au in of order a year. The change in orbit will expose it to an increased probability of a collision, likely resulting in a rapid rise in the infrared flux.  In other words, the region in the inner au will be in turmoil, with objects of various sizes on different trajectories, i.e. rubble from recent collisions on eccentric and inclined orbits and moderate sized grains spiraling rapidly inward toward the star. The result will be a very high rate of collisions and production of debris dust, as observed. The behavior is consistent with expectations for the planetary scattering or migration hypothesis.

\subsection{The Origin of Gas Reservoir }
\label{sec:gas_reservorir}

As mentioned earlier, the presence of circumstellar gas is inferred by the complex, optical line profiles observed in H$\alpha$, Ca II, Na I lines. The H$\alpha$ profiles suggest that the emission structures have velocity widths of $\gtrsim$300 km~s$^{-1}$, originating within a few stellar radii of RZ Psc \citep{punzi18_rzpsc}. Using a magneto-accretion model designed to model gas accretion in T Tauri-like stars, \citet{dmitriev23_rzpsc} model the line profiles taken during the outburst in 2013 and find that the accretion rate is one order of magnitude higher during the outburst, and the magnetosphere extends from $\sim$5 to 10 stellar radii ($\sim$0.03--0.06 au) much larger than typical value for T Tauri-like stars.  It is interesting to note that the derived inclination angle of RZ Psc is 43\arcdeg$\pm$3\arcdeg\  with a relatively weak dipole field strength of $\sim$0.1 kGs. Given the disk inclination angle ($\sim$75\arcdeg), there appears to be a large obliquity in the RZ Psc system. If confirmed, the RZ Psc system would be the sixth solar-like system that exhibits large ($\gtrsim$30\arcdeg) obliquity between the disk and the star as discussed by \citet{hurt_macgregor23}. \citet{dmitriev23_rzpsc} speculate that the extended and weak magnetic field of RZ Psc could result in a long-lived primordial accretion disk. Assuming the circumstellar gas is primordial (i.e., H$_2$ dominated and an ISM-like CO-to-H$_2$ mass ratio of 10$^{-4}$), the mass of the gas reservoir ($\lesssim$10$^{-7} {\rm M_\odot}$, estimated from the ALMA CO upper limit in Section \ref{sec:alma_intep}) could sustain the low accretion rate over 10$^5$ years. Given the age of the system and lack of cold gas as traced by CO observation, there is no clear evidence that the hot gas is primordial as suggested by the magneto-accretion model. 

Tenuous gas has been found in some young debris disks \citep{hughes18}. \citet{punzi18_rzpsc} suggested that the hot gas revealed from the optical spectroscopy could originate from the accretion of material from a hot Jupiter-like planet swallowed by the star. The aftermath of such an engulfment might have been observed recently in ZTF SLRN-2020 \citep{de23_ztf2020}. If true, the detailed evolution of the hot gas in RZ Psc might provide an additional test bed for such a phenomenon. The gas could also originate from out-gassing comets scattered toward the star, collisions of volatile-rich asteroids, or stripped atmospheres from giant impacts between protoplanets. \citet{najita23} recently review these processes and conclude that all of them are capable of generating 10$^{-3}$--10$^{-2} {\rm M_\oplus}$ of secondary gas for stars younger than a few 100 Myr. If the gas is secondary in origin, the ALMA CO gas limit ($<10^{-5} {\rm M_\oplus}$) strongly disfavors the giant impact origin as the estimated mass of stripped atmosphere is on the order of $10^{-4} {\rm M_\oplus}$ for each giant impact \citep{najita23}. It is difficult to differentiate the remaining two mechanisms because both asteroids and comets could store volatile material when they formed, and would later release it in the inner region. However, inward-scattering of comets is a low efficiency process \citep{bonsor12} and often requires a chain of low-mass planets and/or a massive reservoir of icy planetesimals. Even accounting for the entire mm flux, there is no massive reservoir of icy planetesimals (material outside $\sim$10 au) as the potential source of scattered comets. Vigorous collisions between volatile-rich asteroids, consistent with the numerous observed optical dips, is more likely the source of the secondary gas in RZ Psc.

\subsection{The Mechanisms for the Puffed-up Disk Scale Height}
\label{sec:puffed-up}

For gas-poor debris disks, the disk scale height is one of the observables that can constrain a system's dynamical excitation, and hence provide information about the processes and bodies shaping it under the influence of gravity  \citep{wyatt_dent02,quillen07_diskscaleheight,thebault09_verticalstructures}. Similar to gas-rich protoplanetary disks, the disk vertical structure is also strongly stratified in terms of grain sizes -- the larger the grains, the closer to the midplane in gas-poor debris disks. The disk scale height inferred from 
small grains and/or gas species has less diagnostic power in constraining the size of the stirring bodies due to the higher influence of non-gravitational forces such as the radiation pressure and gas drag \citep{thebault09_verticalstructures,olofsson22_vertical_impact_of_gas}.
A disk scale height derived from SED modeling would have different behavior from that described above (not an observed quantity).
As stated in Section \ref{sec:sed_mod}, it is uncertain to pin point the exact number for the disk scale height due to the marginally resolved ALMA image and SED model degeneracy, but a larger than nominal disk scale height appears to be present in the RZ Psc system. The question is what causes it?

The newly discovered, low-mass companion, RZ Psc B, is expected to play an important role in the behavior of the debris system, as discussed by \citet{kennedy20_mnras_496_75_RZpsc_companion}. The Hill radius of the companion is $\sim$10 au assuming a mass of 0.12 M$_\odot$ at a separation of 22 au. The disk outer radius ($\lesssim$13 au derived from the ALMA observation, and 12$^{+1}_{-2}$ au derived from the SED model) is consistent with truncation by the companion. Furthermore, as discussed by \citet{nesvold16}, a stellar-mass perturber orbiting exterior to and inclined to the disk midplane can excite planetesimals into high eccentricity via the Kozai--Lidov mechanism \citep{naoz16review_kozai}, creating debris in significantly extended vertical structures compared to those caused by an interior planet. Based on single epoch data, little is known about the orbital parameters of the companion (except for an estimated mass from its brightness, and projected distance), but it locates at a P.A. of $\sim$75\arcdeg\ that is very different from the disk's P.A. ($-$16\arcdeg$^{+44\arcdeg}_{-20\arcdeg}$) measured by ALMA. It is probable that the relative inclination between the external companion and the disk falls into the allowable range where the eccentric Kozai--Lidov mechanism is active \citep{naoz16review_kozai}. Because the Kozai timescale is inversely proportional to the orbital period of planetesimals, a hallmark for the Kozai perturbation will be highly eccentric particles found in the outermost region closer to the perturber, while particles closer to the star have relatively lower eccentricities. This is in line with the discussion in Section \ref{sec:origin-variability} that the collision activity in the inner au region is less affected by RZ Psc B. 

Nonetheless, the presence of the external companion might provide the initial trigger for the planet migration hypothesis responsible for the inner activity. In this case, the birth place of the migrating planets is likely to be well beyond the snowline and close to the outer region of the disk where the influence of the companion is most effective. Furthermore, the potential misalignment between the companion and the disk midplane would create a potential warp or spiral structure, manifest as a puffed-up disk scale height when viewing close to edge on. In this case, one would expect asymmetric structure between the two disk ansae.  That is, if the intense activity in the inner zone is caused by fully formed giant planets that are in low eccentricity orbits, the debris disk in RZ Psc is expected to be axi-symmetric like many of the Kuiper-belt analogs revealed by ALMA \citep{matra23_reasons} and vertically thin \citep{nesvold16}. If the low-mass companion played a significant role in shaping the disk dynamical activity, we should expect an asymmetric disk \citep{stuber23}. Future higher resolution disk images will provide further evidence to differentiate the two scenarios.

Can planet migration alone create a large disk scale height? The small bodies in our own Kuiper belt have a bimodal distribution (dynamically cold with low inclinations and dynamically hot with high inclinations,   \citealt{brown01}). The bimodal distribution in the small bodies is thought to be a telltale sign of Neptune's migration \citep{nesvorny15_neptunemigration}. Among the exoplanetary systems, $\beta$ Pic is the only one known to show a clear bimodal distribution using both data in optical scattered light (tracing small grains) and millimeter emission (tracing large grains) \citep{golimowski06,matra19b}. The vertical structure along with the disk asymmetry observed in scattered light and the cold CO gas \citep{dent14_betapic} in the $\beta$ Pic system are likely caused by the migration of a yet-to-be discovered super Neptune \citep{matra19b}.

The large vertical scale height in the RZ Psc system is consistent with the suggestion in the preceding section (Section \ref{sec:origin-variability}) that the intense collisional activity in the innermost part of the system arises through the effects of planetary migration. The asteroid belt in our solar system has experienced significant depletion (99\%) compared to its primordial phase \citep{weidenschilling77} and the early depletion was linked to the giant planet migration as suggested by the Grand Tack model \citep{walsh11_grandtack}. If there were giant planets in the RZ Psc system, they must have fully formed already because of the lack of cold gas in the system.  In our solar system, fully formed giant planets played a critical role in shaping the terrestrial planets \citep{raymond14}. In this case, RZ Psc might represent the beginning stage in cleaning its primordial asteroid belt under the influence of newly formed giant planets. In this case, the observed disk scale height is directly linked to the required mass of the scattering planet. If the RZ Psc disk does have an opening angle as large as 50\arcdeg\ (unlikely as we discuss earlier due to the limitation of SED modeling), the planet needs to be, at least if not more than, as massive as Jupiter (with an escape velocity of 60 km~s$^{-1}$) to form a scattered disk at $\sim$5--10 au. Such a planet (Jupiter-like at $<$5 au) should have detectable signal in radial velocity data. Sporadic radial velocity variability was detected by \citet{punzi18_rzpsc}, but the origin is unknown and a future detailed, long-term radial velocity study would provide better constraints on the presence of giant planets in the RZ Psc system.

\section{Conclusion }
\label{sec:conclusion}

We present multiyear infrared monitoring data from {\it Spitzer} and {\it WISE} to track the activities of inner debris production/destruction in RZ Psc. Millimeter observations along with the SED modeling provide a good assessment of the overall disk properties. These observations confirm that RZ Psc hosts a close to edge-on, highly perturbed disk, consistent with numerous optical dips caused by dust clumps in the past five decades. 

ALMA 1.3 mm data obtained in 2021 reveal a compact, marginally resolved disk with a synthesized beam of 0\farcs06$\times$0\farcs04. Parametric models in fitting the visibilities suggest the millimeter emission has a radial extent less than 13 au, inclined by 83\arcdeg$^{+13\arcdeg}_{-24\arcdeg}$ from face on, with a P.A. of $-$16\arcdeg$^{+44\arcdeg}_{-20\arcdeg}$. The total 1.3 mm flux is 42$\pm$8 $\mu$Jy, corresponding to a dust mass of 6.4--40 $\times 10^{-3} {\rm M_\oplus}$. No cold CO(2-1) line was detected, suggesting $<$1.0$\times 10^{-5} {\rm M_\oplus}$ for the CO gas mass. The ALMA data indicate a very low gas-to-dust mass ratio $<$$10^{-3}$, confirming that RZ Psc hosts a gas-poor debris disk and has evolved out of the gas-rich protoplanetary disk stage. Adopting an axe-symmetric disk geometry, SED models suggest the disk inner radius is near $\sim$0.1 au, close to the dust sublimation radius, in order to produce enough infrared excess in the 3--5 $\mu$m region. The disk outer radius is at 12$^{+1}_{-2}$ au (consistent with the derived ALMA value) with a large scale height, suggesting a wide opening angle of $\sim$50\arcdeg. Although the disk scale height and inclination are degenerate in SED models, the detected percentage of the optical polarization ($\sim$1\%) further suggests that the disk is inclined by $\sim$75\arcdeg, consistent with the ALMA derived value and that a large disk scale height is necessary to fit the prominent 10 $\mu$m feature. The SED derived millimeter flux is 3 times lower than the ALMA value, suggesting the disk might have a more complex structure than axi-symmetric. An additional cold component that only emits at far-infrared or millimeter wavelengths, or an eccentric disk where grains extend further out radially both can account for the missing flux in the SED. 

The four-year, high-cadence ($\sim$3--5 days) 3--5 $\mu$m data reveal that the RZ Psc system exhibits stochastic weekly infrared variability with a typical flux change of $\pm$5--7 mJy at 4.5 $\mu$m over a year if not more. Adopting a dust temperature of $\sim$1000 K under the optically thin condition, the 4.5 $\mu$m flux variation corresponds to a minimum change in the dust surface area of $\sim$5$\times10^{-4}$ au$^2$ or 10$^{21}$ g. The intense collisions in the inner planetary zone totally destroy a mass equivalent to a 90-km-size asteroid on yearly basis. If such a high rate has persisted over a period of 10$^3$ (10$^6$) years, a Moon- (Earth-)size object would have been destroyed. 

Although there is no cold CO gas detected in the RZ Psc system, the complex, variable optical line profiles do suggest the presence of close-in circumstellar gas with a very low accretion rate (10$^{-12} M_\odot$ yr$^{-1}$) if assuming an H2-dominated, primordial disk. Alternatively, the hot gas could be the remnant of a Jupiter-like planet engulfment as suggested by \citet{punzi18_rzpsc}. Furthermore, the gas could also be secondarily created by out-gassing comets, collisions of volatile-rich asteroids, or stripped atmospheres from giant impacts between protoplanets \citep{najita23}. Collisions between volatile-rich asteroids are likely the source of the secondary gas given that (1) the low efficiency in inward scattering of comets and that  RZ Psc does not have a massive cold reservoir of Kuiper-belt zone outside $\sim$10 au, where Jupiter-family comets in our solar system originate, and (2) stripped planetary atmospheres are expected to release more gas than observed. 

It is evident that RZ Psc hosts a debris disk highly perturbed either by recently formed giant planets and/or by its low-mass companion RZ Psc B. The intense collision activity in the inner au region is likely a result of planetary migration or planet-planet scattering, both deflecting large numbers of planetesimals onto colliding orbits and resulting in large amounts of collisionally produced dust manifest as stochastic infrared variability and numerous optical dips. RZ Psc might be in the beginning stage of giant planet migration that has been initiated by gravitational effects caused by the companion if the migrating planets were formed beyond the snowline. Furthermore, the large disk scale height might also be related to the low-mass companion. In this case, one would expect the disk to show warp or spiral structures induced by the external perturber similar to the ones found around HD 106906 \citep{fehr22_hd106906}. Follow-up observations to better determine the orbital parameters of the companion, and future high-resolution disk images will shed more light into the nature of the highly perturbed RZ Psc disk.

\appendix 
\restartappendixnumbering

\section{Warm {\it Spitzer} Photometry for RZ Psc}

\begin{deluxetable*}{ccrrrrcrrrr}
\tablewidth{0pc}
\footnotesize
\tablecaption{The IRAC fluxes of the RZ Psc system\label{tab:irac}}
\tablehead{
\colhead{AOR Key}&\colhead{BMJD$_{3.6}$}&\colhead{$F_{3.6}$}&\colhead{$E_{3.6}$}&\colhead{$F_{IRE,3.6}$}&\colhead{$E_{IRE,3.6}$}  &\colhead{BMJD$_{4.5}$}& \colhead{$F_{4.5}$}&\colhead{$E_{4.5}$}&\colhead{$F_{IRE,4.5}$}&\colhead{$E_{IRE,4.5}$}    \\ 
\colhead{  }&\colhead{(day) }&\colhead{(mJy)}&\colhead{(mJy)}&\colhead{(mJy)}&\colhead{(mJy)}&\colhead{(day)}&\colhead{(mJy)}&\colhead{(mJy)}&\colhead{(mJy)}&\colhead{(mJy)}
}
\startdata
 53391872 &    57096.86790 &    57.63 &     0.22 &    23.63 &     0.55 &    57096.86548 &    50.18 &     0.18 &    28.18 &     0.38  \\ 
  53391360 &    57101.24285 &    56.91 &     0.46 &    22.91 &     0.68 &    57101.24045 &    50.76 &     0.18 &    28.76 &     0.38  \\ 
  53391104 &    57104.72590 &    58.13 &     0.27 &    24.13 &     0.57 &    57104.72351 &    51.63 &     0.15 &    29.63 &     0.36  \\ 
  53390848 &    57108.53130 &    58.07 &     0.29 &    24.07 &     0.59 &    57108.52895 &    51.21 &     0.13 &    29.21 &     0.36  \\ 
  53390592 &    57112.90097 &    53.18 &     0.27 &    19.18 &     0.57 &    57112.89862 &    45.74 &     0.14 &    23.74 &     0.36  
\enddata
\tablecomments{$F$ and $E$ are the flux and uncertainty including the star, while $F_{IRE}$ and $E_{IRE}$ are the excess quantities excluding the star.  This table is published in its entirety in the
machine-readable format. A portion is shown here for guidance regarding its form and content.}
\end{deluxetable*}

Table \ref{tab:irac} shows the 3.6 and 4.5 $\mu$m photometry of the RZ Psc system obtained during the {\it Spitzer} warm mission as described in Section \ref{sec:irac_obs}. The excess emission and its uncertainty were derived by the subtraction of the expected photospheric value (34 and 22 mJy at the 3.6 and 4.5 $\mu$m bands, respectively) with a typical uncertainty of 1.5\% of that value added in quadrature.

\facilities{{Spitzer} (IRAC), WISE, ALMA, IRSA, AAVSO}

\begin{acknowledgments}

We thank the referee for providing constructive comments that improved the clarity of this manuscript.
This work has been supported by NASA ADAP programs (grant No. NNX17AF03G and 80NSSC20K1002). H.B.L. is supported by the National Science and Technology Council (NSTC) of Taiwan (Grant Nos. 111-2112-M-110-022-MY3). 
The paper is based on observations made with the {\it Spitzer} Space Telescope, which was operated by the Jet Propulsion Laboratory, California Institute of Technology. This paper has made use of data from AAVSO, and we acknowledge with thanks the variable star observations from
the AAVSO International Database contributed by observers worldwide and used in this research.
The MCMC part of the work was supported by the University of Arizona High-Performance Computing (HPC) resources. 
This paper makes use of the following ALMA data: ADS/JAO.ALMA\#2019.1.01701.S. ALMA is a partnership of ESO (representing its member states), NSF (USA) and NINS (Japan), together with NRC (Canada), MOST and ASIAA (Taiwan), and KASI (Republic of Korea), in cooperation with the Republic of Chile. The Joint ALMA Observatory is operated by ESO, AUI/NRAO and NAOJ. 

This research has made use of the NASA/IPAC Infrared Science Archive, which is funded by the National Aeronautics and Space Administration and operated by the California Institute of Technology. 
This work has also made use of data from the European Space Agency (ESA) mission{\it Gaia} (\url{https://www.cosmos.esa.int/gaia}), processed by the {\it Gaia} Data Processing and Analysis Consortium (DPAC, \url{https://www.cosmos.esa.int/web/gaia/dpac/consortium}). Funding for the DPAC has been provided by national institutions, in particular the institutions participating in the {\it Gaia} Multilateral Agreement. 

The Submillimeter Array is a joint project between the Smithsonian
Astrophysical Observatory and the Academia Sinica Institute of Astronomy
and Astrophysics, and is funded by the Smithsonian Institution and the
Academia Sinica.


\end{acknowledgments}

\bibliography{ms.bbl}

\end{document}